\documentclass[preprint,12pt,authoryear]{elsarticle}

\usepackage{hyperref}
\usepackage{astrojournals}
\usepackage{amsmath}
\usepackage{graphicx}
\usepackage{txfonts}
\usepackage{color}
\usepackage[percent]{overpic}
\AtBeginDocument{\mathcode`v=\varv}

\newcommand{\noted}[1]{\textcolor{red}{#1}}

\journal{New Astronomy}

\begin{document}

\begin{frontmatter}

  \title{Emission from elliptical streams of dusty debris around white dwarfs}

  \author[label1]{C.~J.~Nixon\corref{cor1}} \ead{cjn@leicester.ac.uk}
  \author[label1,label2]{J.~E.~Pringle}
  \author[label3,label4]{E.~R.~Coughlin}
  \author[label5]{A.~Swan}
  \author[label5]{J.~Farihi}
  \cortext[cor1]{Corresponding author}
  \address[label1]{School of Physics and Astronomy, University of Leicester, Leicester, LE1 7RH, UK}
  \address[label2]{Institute of Astronomy, Madingley Road, Cambridge, CB3 0HA, UK}
  \address[label3]{Department of Astrophysical Sciences, Princeton University, Princeton, NJ 08544, USA}
  \address[label4]{Department of Physics, Syracuse University, Syracuse, NY 13244, USA}
  \address[label5]{Department of Physics and Astronomy, University College London, Gower Street, London WC1E 6BT, UK}

\begin{abstract}
White dwarfs are routinely observed to have polluted atmospheres, and sometimes significant infrared excesses, that indicate ongoing accretion of circumstellar dust and rocky debris. Typically this debris is assumed to be in the form of a (circular) disc, and to originate from asteroids that passed close enough to the white dwarf to be pulled apart by tides. However, theoretical considerations suggest that the circularisation of the debris, which initially occupies highly eccentric orbits, is very slow. We therefore hypothesise that the observations may be readily explained by the debris remaining on highly eccentric orbits, and we explore the properties of such debris. For the generic case of an asteroid originating at several au from the white dwarf, we find that all of the tidal debris is always bound to the white dwarf and that the orbital energy distribution of the debris is narrow enough that it executes similar elliptical orbits with only a narrow spread. Assuming that the tidal field of the white dwarf is sufficient to minimise the effects of self-gravity and collisions within the debris, we estimate the time over which the debris spreads into a single elliptical ring, and we generate toy spectra and lightcurves from the initial disruption to late times when the debris distribution is essentially time steady. Finally we speculate on the connection between these simple considerations and the observed properties of these systems, and on additional physical processes that may change this simple picture.
\end{abstract}

\begin{keyword}
  circumstellar matter \sep minor planets, asteroids: general \sep planetary systems \sep white dwarfs
\end{keyword}

\end{frontmatter}

\section{Introduction}
White dwarfs are the end state of stars with masses less than about $9M_\odot$ \citep[e.g.][]{Shapiro:1983aa}. The masses of known white dwarfs spans $\approx 0.1M_\odot$ to $1.4M_\odot$ with a strong peak towards $0.6M_\odot$ \citep[e.g.][]{Kepler:2007aa}. White dwarfs are initially born with temperatures $\gtrsim 10^5$\,K, and over a period of $\sim 20$\,Myr they cool to a temperature of $\approx 25,000$\,K. At cooling ages of $\sim 30-600$\,Myr, white dwarfs can be observed to have polluted atmospheres indicated by the presence of atmospheric metal absorption lines in their spectra and infrared spectral excesses thought to indicate the presence of dust heated to temperatures of order a few hundred to $\sim 10^3$\,K \citep[see e.g.][for discussion]{Farihi:2016aa}. 

The fraction of all white dwarfs that show evidence for continued accretion of planetary material in the form of polluted atmospheres or infrared excesses is large \citep[$25-50$ per cent; e.g.][]{Zuckerman:2003aa,Zuckerman:2010aa,Koester:2014aa}. The currently accepted explanation for the origin of this material is that it stems from a reservoir of asteroid-like objects that survived the red giant phase of the white dwarf's progenitor star. Such a reservoir must therefore be at radii larger than $\approx 5-10$\,au, and a mechanism is needed to scatter these objects inwards towards the white dwarf. \cite{Debes:2002aa} proposed that the extensive mass loss that takes place during the red giant phase leads to rearrangement of the outer planetary orbits, and thereby destabilisation of the orbits of any reservoir of asteroid-like objects. \cite{Bonsor:2011aa} consider scattering in objects from a Kuiper belt by a Neptune mass planet set at 30 au. They find that (given enough objects as deduced from an observed planetesimal-belt population on the main-sequence) they can scatter objects inwards but not far enough. \cite{Frewen:2014aa} showed that a planet in an eccentric orbit at a radius of around 4 au is capable of scattering material further at the requisite rate.

Rather than impacting directly onto the white dwarf, it is more likely that an asteroid falling in from several au will pass within the tidal disruption radius (Roche radius) and be torn apart. Such a process is already well discussed both in the context of the solar system (e.g. cf. Comet Shoemaker-Levy 9, \citealt{Chapman:1993aa}) and with regard to the tidal disruption of stars by massive black holes (known as TDEs, e.g. \citealt{Rees:1988aa}). For asteroids, the tidal radius, $R_{\rm t}$, is given approximately by \citep{Bear:2013aa}
\begin{equation}
  \label{rt}
  R_{\rm t} \approx 1.3 \left( \frac{\rho_{\rm ast}}{3 {\rm g/cm}^3} \right)^{-1/3} \left( \frac{M_{\rm wd}}{0.6 M_\odot} \right)^{1/3} R_\odot,
\end{equation}
where $\rho_{\rm ast}$ is the (assumed to be constant) density of the asteroid and $M_{\rm wd}$ is the mass of the white dwarf.

Once such a disruption has taken place it is generally assumed \citep{Jura:2003aa} that the dusty debris settles down in the form of a disc. This disc, being heated by the central white dwarf, then provides a source of the observed infrared excesses \citep{Jura:2003aa,Reach:2009aa} and for the accretion of material into the white dwarf atmosphere \citep[e.g.][]{Rafikov:2011aa,Rafikov:2011ab}. A major problem with this picture is that the stream of disrupted debris lies initially on highly eccentric orbits, and it is not at all clear how dust on such orbits manages to lose enough energy to become part of a circular debris disc. Effects such as general relativistic precession from the white dwarf's gravity and Poynting-Robertson drag from the white dwarf radiation field have all been considered and are generally found to affect the debris orbits only on very long timescales (we provide discussion and timescales for these effects in Section~\ref{mult_ast}).

Focussing on the systems that display a clear excess infrared flux (not attributed to a stellar or substellar companion), signaling the presence of a significant amount of dust, then from an observational standpoint there is little direct evidence to motivate the {\it assumption} that the dust orbits in a (near) circular disc (as is assumed in the models of e.g. \citealt{Rafikov:2011aa,Rafikov:2011ab}). What is seen \citep[see e.g.][for a review]{Farihi:2016aa} is typically excess flux at near-infrared to mid-infrared wavelengths, say $\approx 2-20\,\mu$m. At wavelengths shorter than $\approx 2\,\mu$m, the spectra are typically dominated by the emission from the white dwarf. At wavelengths longer than several tens of microns observations show mostly upper limits to the fluxes. At about $8-12$\,$\mu$m there is a strong silicate feature revealed by {\it Spitzer}/IRS spectra. These observations indicate that rocky debris exists at radii close to the white dwarf (of order the tidal and/or sublimation radius), but they do not provide any information about the orbits that the debris occupies. For example, while a model of dust distributed in a disc can fit the observed infrared excess \citep{Jura:2007aa,Farihi:2009aa}, so can a model composed of an optically thin `shell' of material (see e.g. \citealt{Reach:2005aa,Reach:2009aa} who find such a model can fit both the continuum and emission components of G29-38). \cite{Farihi:2016aa} argues that such (non-disc) models are usually ignored as the mass constraint of optically thin discs ($\lesssim 10^{18}$\,g) is too low to explain the accretion histories ($\gtrsim 10^{23}$\,g in many systems). However, debris in the form of one, or several, ellipses may be able to contain the required mass reservoir.

To understand the orbital properties of a system it is necessary to turn to its variability properties. There are now several variability studies relevant to the excess infrared emission from rocky debris that we summarise here, but we note that the cadence of such observations is often too poor (with observations of the same system typically separated by timescales of order years) to draw concrete conclusions. Starting with studies looking at one or two sources, we have: (1) \cite{Xu:2014aa} report a drop in the infrared luminosity of J0959--0200 observed with {\it Spitzer}/IRAC by about 35 per cent on a timescale of less than 300 days. The drop occurred in the flux at both 3.6\,$\mu$m and 4.5\,$\mu$m (a weaker drop was also seen in the K band). (2) \cite{Xu:2018aa} report infrared variability of two systems J1228+1040 and G29-38: J1228+1040 is known to host a gas disc \citep[e.g.][]{Manser:2016aa}. This system showed a drop in flux at 3.6\,$\mu$m and 4.5\,$\mu$m of 20 per cent between 2007 and 2014, and remained at this level when observed again in 2018. For G29-38, the emission from the silicate feature at 10\,$\mu$m increased by 10 per cent between 2004 and 2007. (3) \cite{Farihi:2018aa} report a long duration lightcurve of GD 56 at wavelengths of $\approx 3.5\,\mu$m and $4.5\,\mu$m (both {\it Spitzer} and {\it WISE} data) that show a rise and fall by 20 per cent over 11.2 years. The ratio of the fluxes between the two bands is approximately constant over the duration of the observations in both the {\it Spitzer} and {\it WISE} data. (4) \cite{Wang:2019aa} show an increase in flux in both {\it WISE} bands 1 \& 2 by one magnitude within half a year, and the flux continued at this level for another year or so. There have also been some population studies: (5) \cite{Swan:2019aa} use archival data from {\it WISE} to find infrared flux variations of tens of per cent around approximately half of dusty white dwarfs in their sample. For the WD 0956--017 they discuss that the flux at infrared wavelengths decreased by 35 per cent during 2010 and was still low in 2014 \citep{Xu:2014aa}, but the flux increased after 2014 by a factor of almost two during NEOWISE. \cite{Swan:2019aa} primarily used data in the $3.4\,\mu$m band over a period of seven years. These authors point out that such variability is at odds with the standard disc model. (6) Rogers et al. (submitted) explore the near infrared variability of 34 dusty white dwarfs with J, H and K band data from UKIRT over a period of 3 years. They find that there is no statistically significant variation in these bands for these objects on this timescale. Thus, in summary, it appears that white dwarfs with excess infrared flux routinely show variability at mid-infrared wavelengths with amplitudes of tens of per cent and timescales of months to years, and occasionally they show large amplitude (of order 100 per cent) changes on similar timescales. At shorter wavelengths, where the white dwarf dominates the emission, there is little or no variability. This picture is not complete with almost no data with high cadence (e.g. on timescales less than days to weeks), and little is known about the variability at wavelengths $\gtrsim 10\,\mu$m. 

In this paper, we propose and explore the possibility that most of the dust remains in the form of an elliptical stream, or perhaps multiple such streams if the lifetime of the streams exceeds the timescale on which further asteroids are scattered inwards. In Section \ref{simplestream} we explore a simple model where the debris stream is in the form of a single ellipse with mass distribution determined by a constant mass flux around the ellipse. We also provide a simple estimate of the spectrum resulting from this ring. In Section~\ref{filling} we explore the period of time between the disruption of the asteroid and the filling of the ellipse with debris. We calculate the spectra that result through this process and provide an example lightcurve. We discuss our results, their caveats, and some additional physical processes that may be important in Section~\ref{discussion}. Finally, in Section~\ref{conclusions} we provide our conclusions.

\section{The simple steady-state debris ellipse}
\label{simplestream}
The picture we have in mind, following \cite{Debes:2002aa} and \cite{Frewen:2014aa}, is a central white dwarf with a planetary perturber at several au and an asteroid belt near the perturber. Every now and again an asteroid is kicked in on a highly eccentric orbit that passes close to the white dwarf. For asteroid orbits that pass inside the tidal sphere of the white dwarf (equation~\ref{rt}), these asteroids will be destroyed and the resulting debris forms a narrow stream as the tidal field stretches the asteroid in one direction while compressing it in the other two \citep[e.g.][]{Coughlin:2016ab,Coughlin:2016aa}. In this section, we show that the spread of orbits imparted to the asteroid debris leads to the formation of a debris ellipse, and we provide an estimate of the timescale on which this occurs and provide a simple model of the spectrum arising from such an ellipse.

\subsection{Formation of the debris ellipse}
The general behaviour of the debris following the tidal disruption of an object, whether it be an asteroid disrupted by a white dwarf or a star disrupted by a supermassive black hole, can be estimated by considering the relevant physics \citep[cf.][]{Rees:1988aa}.

If we assume that the asteroid is scattered inwards from an apastron distance of $R_{\rm a} = 4$\,au, then it approaches the white dwarf on a highly eccentric orbit with a semi-major axis, $a \approx 2$\,au. If the periastron distance, $R_{\rm p}$, is equal to the tidal radius (equation~\ref{rt}), then the semi-major axis of the orbit is $a = (R_{\rm a} + R_{\rm p})/2 = 2.003\,{\rm au}$ and the eccentricity is $e = (R_{\rm a}-R_{\rm p})/(R_{\rm a}+R_{\rm p}) = 0.997$. The orbital period, $P$, is given by 
\begin{equation}
  \label{P}
P = 2\pi\sqrt{\frac{a^3}{GM_{\rm wd}}} = 3.66 \left( \frac{a}{2 {\rm au}} \right)^{3/2}  \left( \frac{M_{\rm wd}}{0.6 M_\odot} \right)^{-1/2} {\rm yr}\,.
\end{equation}
The energy (per unit mass) of this orbit is given by 
\begin{equation}
  \label{E0}
  E_0 = - \frac{G M_{\rm wd}}{2a} = - 1.33\times10^{12}\left( \frac{M_{\rm wd}}{0.6 M_\odot} \right) \left( \frac{a}{2 {\rm au}} \right)^{-1} {\rm erg}\,.
\end{equation}

To a first approximation the average orbital energy of the disrupted debris is  that of the centre of mass of the asteroid, i.e. $\approx E_0$. However, the ``frozen-in'' approximation \citep[e.g.][]{Lacy:1982aa,Lodato:2009aa,Stone:2013aa,Coughlin:2019aa} gives that the orbits of the debris have a spread of energies $E_0 - \Delta E \le E_0 \le E_0 + \Delta E$, where $\Delta E$ is given by
\begin{equation}
  \label{deltaE}
\Delta E \approx k\frac{G M_{\rm wd} R_{\rm ast}}{R_{\rm t}^2}\,,
\end{equation}
where $R_{\rm ast}$ is the radius of the asteroid, and $k$ is a number of order unity that depends on the structure and properties of the object being disrupted. This equation is valid for all orbits where the asteroid passes within the tidal sphere of the white dwarf. If we define the impact parameter $\beta \equiv R_{\rm t}/R_{\rm p}$, then the minimum $\beta$ that leads to disruption is $\approx 1$ and the maximum $\beta$ to avoid direct impact of the white dwarf surface ($R_{\rm p} = R_{\rm wd}$) is $\beta \approx 130$. For any value of $\beta$ in this range, the energy spread is imparted to the debris at the moment the asteroid is disrupted. This occurs when the asteroid {\it enters} the tidal disruption sphere, and therefore the energy spread is set by the tidal shear at the tidal radius and is given by equation~\ref{deltaE} \citep{Lacy:1982aa,Stone:2013aa}, with $k$ expected to be within a factor of a few from unity \citep[e.g.][]{Steinberg:2019aa}.\footnote{It is possible that a combination of effects, such as collisions and self-gravity of the debris, can lead to alterations of the energy spread (cf. e.g. \citealt{Coughlin:2016aa} who explore this possibility for the stellar disruption case).}

For the parameters discussed above, and assuming $k=1$, we find that the energy spread is
\begin{equation}
\Delta E \approx  1.94\times10^{10} \left( \frac{M_{\rm ast}}{10^{20} {\rm g}} \right)^{1/3} \left( \frac{ \rho_{\rm ast}}{3 {\rm g/cm}^3 } \right)^{1/3} \left( \frac{M_{\rm wd}}{0.6 M_\odot} \right)^{1/3} {\rm erg}\,.
\end{equation}
To increase $\Delta E$ to the value of $|E_0|$ given in equation~\ref{E0}, and thus generate parabolic and hyperbolic orbits of the debris, requires an increase in the product of the density and mass of the disrupted body $\rho M \sim 10^6$. As the density of rocky bodies in the solar system are typically found to have a density of order a few g/cm$^3$, this implies that (for apastron radii of order a few au) the mass of the disrupted body must be $\gtrsim 10^{26}$\,g (e.g. the size of a small planet, say Mercury, rather than of asteroid size). The rate of tidal disruption of large bodies is much smaller than the rate of disruption of small bodies as the smaller bodies are more numerous and more easily gravitationally perturbed.

Hence we find that in general $\Delta E/|E_0| \ll 1$, and therefore conclude that the debris will be initially in the form of a narrow stream. This spread in energies leads to a spread in orbital periods for the debris given by\footnote{Note that we have $2\Delta E$ in this equation because the width of the energy spread around $E_0$ is both plus and minus $\Delta E$.} $\Delta P/P \approx (3/2) (2\Delta E/(-E_0))$. Therefore, neglecting interactions among the debris, the debris fills the eccentric orbit after a time, $t_{\rm fill} \approx P^2/\Delta P$, given by\footnote{We note that the concept of a filling timescale is not applicable to debris with $\Delta E \gtrsim |E_0|$, which occurs routinely in the case of stellar disruption by a supermassive black hole where the orbit is essentially parabolic and $E_0 \approx 0$, as the orbital period of parabolic or hyperbolic orbits is not well-defined. We point the interested reader to \cite{Veras:2014aa}, who provide formulae for the filling time for debris with parabolic/hyperbolic orbits.}
\begin{equation}
  \label{tfill}
t_{\rm fill} \approx 83.4 \left(\frac{a}{2 {\rm au}} \right)^{1/2}  \left( \frac{M_{\rm wd}}{0.6 M_\odot} \right)^{1/6} \left( \frac{M_{\rm ast}}{10^{20} {\rm g}} \right)^{-1/3} \left( \frac{ \rho_{\rm ast}}{3 {\rm g/cm}^3 } \right)^{-1/3} {\rm yr}\,.
\end{equation}

The general properties of the above discussion is that (1) the energy distribution imparted to the debris is narrow and in particular much smaller than the magnitude of the energy of the original asteroid orbit -- this ensures that all of the debris remains bound to the white dwarf, and (2) the timescale on which the debris fills out the ellipse is many orbits (approx. 25 orbits for the above parameters) but much shorter than evolutionary timescales for the system. These findings agree with existing works in the literature that simulate the disruption of rubble pile asteroids including the effects of, e.g., the debris gravity and collisions \citep{Debes:2012aa,Veras:2014aa}\footnote{\cite{Veras:2014aa} also provide an analytical description for the filling time of an ellipse. We employ the same assumptions as \cite{Veras:2014aa} and recover an answer that is entirely consistent. However, our approach employs the basic physics of tidal disruptions \citep[cf.][]{Lacy:1982aa,Rees:1988aa} to significantly simplify the discussion.}. In the next subsection we estimate the spectrum from such an ellipse of debris.

\subsection{Stream spectrum}
It is evident from the above discussion that the most usual spatial configuration for the debris from a disrupted asteroid is in the form of a narrow elliptical stream around the central white dwarf. As shown above, to a zeroth order approximation the disrupted debris fills out the orbit in a fairly short timescale, which is $\sim$100\,yr for a typical asteroid mass of $\sim$10$^{20}$\,g. Given this, we are now in a position, using the simplest assumptions, to calculate a zeroth order approximation to the spectrum emitted by the debris, which we assume is mostly in the form of dust.

The dust occupies an orbit (in cylindrical polar coordinates $\left(R, \theta\right)$) of the form
\begin{equation}
R = \frac{a(1-e^2)}{1 + e \cos \theta},
\end{equation}
which is an ellipse with semi-major axis $a$ and eccentricity $e < 1$. The periastron occurs at $\theta = 0$, at a radius that we expect to be approximately the tidal radius $R_{\rm t} \approx a (1-e)$. The apastron is expected to occur approximately at the radius of the last scatterer, $R \approx 2a$, since $e \approx 1$. 

The velocity, $v$,  of material at a point in the orbit is given by
\begin{equation}
v^2 = \left(\frac{G M_{\rm wd}}{a} \right) \left( \frac{1 + 2 e \cos \theta + e^2}{1 - e^2}  \right).
\end{equation}
At late times ($t \gg t_{\rm fill}$) the debris is in an approximate steady state where the mass flow rate through any point of the orbit is independent of time, and thus we expect the mass of dust per unit length, $\mu(\theta)$, along the orbit to be of the form
\begin{equation}
\label{masstheta}
 \mu(\theta) \propto 1/v(\theta).
\end{equation}
This implies that more material is located towards apastron of the orbit, i.e. where the velocity is slower.

In addition, if the dust is optically thin and radiates locally as a black body, we expect the temperature of the dust at radius $R$ to be
\begin{equation}
  \label{tdust}
T_{\rm dust} = T_{\rm wd} \left( \frac{R_{\rm wd}}{R} \right)^{1/2},
\end{equation}
where $T_{\rm wd}$ is the effective temperature of the white dwarf, and $R_{\rm wd}$ its radius. We should note that things are presumably more complicated than this (parts of the stream may be optically thick, the dust may not emit like a black body, the emitted spectrum may depend on the albedo and mass distribution of the emitting particles) but we ignore these complications in this initial work.

\begin{figure}
  \includegraphics[width=\columnwidth]{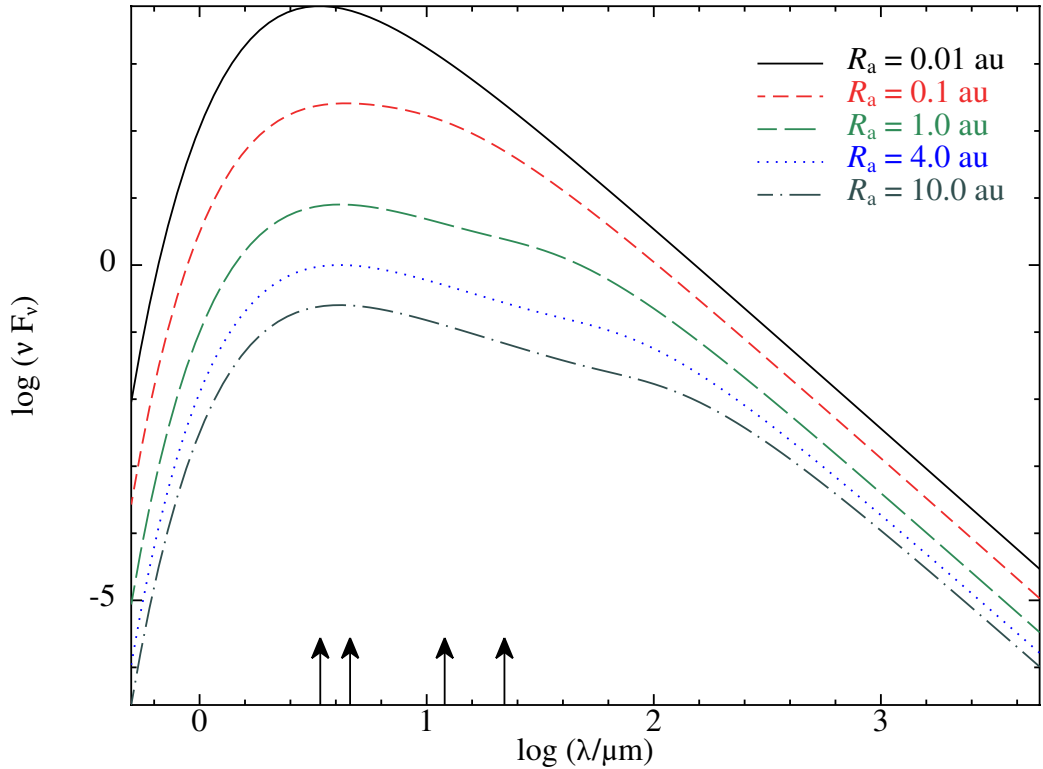}
  \caption{Spectra from the steady state elliptical debris streams produced from the tidal disruption of an asteroid by a white dwarf. The different lines correspond to asteroids that originate from different apastron distances, $R_{\rm a}$, where the black solid line has $R_{\rm a}=0.01$\,au, the red dashed line has $R_{\rm a}=0.1$\,au, the green long-dashed line has $R_{\rm a}=1$\,au, the blue dotted line has $R_{\rm a}=4$\,au, and the grey dot-dashed line has $R_{\rm a}=10$\,au. The fluxes have been scaled to the peak flux for the (fiducial) 4\,au elliptical stream, and the emitting area (i.e. mass and size distribution of the debris) in the same in each model -- thus the smaller ellipses emit more flux due to their higher average temperatures. The effective wavelengths of the {\it WISE} bands (3.4, 4.6, 12.0, and 22.0\,$\mu$m) are marked with black arrows for reference. The figure shows that when the apastron distance ($R_{\rm a}$) is comparable with the periastron distance ($R_{\rm p} = R_{\rm t} = 1.3 R_\odot = 6.1\times 10^{-3}$\,au) the spectra are well-described by single temperature black bodies. This occurs because, in this case, the ratio $T_{\rm max}/T_{\rm min}$ is not large (cf. equation~\ref{tdust}). As the apastron is increased the ratio $T_{\rm max}/T_{\rm min}$ increases, and thus the spectra exhibit a larger fraction of their flux at longer wavelengths (from a few to $\sim$100\,$\mu$m).}
  \label{Fig1}
\end{figure}

By making these straightforward assumptions, we can compute the spectrum of the elliptical stream of dust. The spectrum depends only on two parameters, $T_{\rm max}$ which is the dust temperature at periastron, and $T_{\rm min}$ which is the dust temperature at apastron. If we have $T_{\rm min} \sim T_{\rm max}$ (i.e. a near circular orbit) then we expect a spectrum that is essentially a single temperature black body spectrum. If instead, $T_{\rm min} \ll T_{\rm max}$ (i.e. a highly eccentric orbit) then the spectrum will be a sum of black bodies with a spread of temperatures, which is expected to exhibit a plateau or power-law. To illustrate this, we plot in Fig.~\ref{Fig1} example spectra for elliptical streams of dust. To calculate the spectra we integrate the flux emitted around the orbit assuming that the emitting area is proportional to the mass at each point on the orbit (equation~\ref{masstheta}) and that the temperature is given by equation~\ref{tdust}. The parameters we used are: $M_{\rm wd} = 0.6M_\odot$, $R_{\rm wd} = 0.01R_\odot$, $T_{\rm wd} = 14,000$\,K, $R_{\rm p} = 1.3 R_\odot$, and to illuminate the effect of varying $T_{\rm max}/T_{\rm min}$ we vary the apastron radius between $0.01$ and $10$\,au. We keep the emitting area constant in each case, that is to say that there is the same amount of debris (mass and size) in each stream, and the debris is spread out around the orbit following equation (\ref{masstheta}). Fig.~\ref{Fig1} shows that as the apastron radius is reduced the spectrum approaches a single-temperature blackbody. For example, with $r_{\rm a} = 0.01$\,au, then we have $r_{\rm a}/r_{\rm p} \approx 1.7$ and thus $T_{\rm max}/T_{\rm min} \approx 1.3$. For larger apastra the spectrum contains additional flux at longer wavelengths from the cooler material at large radii, where for example with $r_{\rm a} = 4$\,au we have $r_{\rm a}/r_{\rm p} \approx 660$ and thus $T_{\rm max}/T_{\rm min} \approx 26$. For ellipses of size $R_{\rm a}\gtrsim 1$\,au, the spectrum takes on a characteristic shape with an exponential drop at short wavelengths ($\lambda \lesssim$ a few $\mu$m), a Rayleigh-Jeans tail at $\lambda \gtrsim 100\,\mu$m, and approximately a power-law in between. The power-law is $\propto \lambda^{-\alpha}$ with $\alpha \approx 0.8-0.9$. This is distinct from the classical power law slope of (active and passive) thin discs where typically $\alpha = 4/3$.

Modelling of the infrared excess in the observed spectral energy distributions of white dwarfs generally assumes that the dust is only observed over a narrow range of temperatures, e.g. Fig.~5 of \cite{Farihi:2016aa} shows that $T_{\rm max}/T_{\rm min}\approx 2$ provides a reasonable fit to the infrared excess observed in GD 16 and LTT 8452, and in general a single temperature black body is also sufficient for a reasonable fit (see also \citealt{Dennihy:2017aa}). {\it If} only a narrow range of temperatures is allowed by the observed spectral shape of the infrared excess, then for the elliptical debris stream model this would imply that the dust in the stream is only seen over a narrow range of radii. This, at first glance, appears to suggest that the apastron of the asteroid orbit cannot be too large, which would be inconsistent with the expectation that objects at radii $\lesssim$ a few au would be lost during the red giant phase (and thus the asteroids must originate from radii of several au or larger). However, there may be other explanations such as non-uniformity around the orbit (see next section) or more efficient grinding of larger bodies (that emit less efficiently per unit mass) into dust near periastron where the debris density is highest (see discussion). However, the available data to which models have been fit do not provide a strong constraint on the flux at wavelengths $\gtrsim 10\,\mu$m -- the data here are mostly upper limits and generally have larger error bars than the shorter wavelength data. Thus it is possible that a wider range of temperatures may provide an equally acceptable fit.

\section{The evolution of a more realistic debris stream}
\label{filling}
At late times, $t \gg t_{\rm fill}$, we have seen that our simple assumptions in the previous section lead the eccentric orbit to be uniformly populated with debris (equation \ref{masstheta}). However, for time $t < t_{\rm fill}$, the orbit is only partially populated, and this would lead to time variability on the orbital period $P$. In the first few orbits after disruption of the asteroid this variability will be large amplitude (up to 100 per cent) as the debris moves from periastron to apastron and thus from high to low temperatures. As time increases more of the orbit is filled and thus the duty cycle of the emission increases until after a time $\approx t_{\rm fill}$ when the duty cycle is 100 per cent with a significant flux across the full orbital period. As time increases there is overlap between the ends of the stream and therefore the density of material around the orbit varies, with the overlapping regions containing more debris. After a time post disruption of around $n < t/t_{\rm fill} < (n+1)$ we expect the overdensity to be of order $(n+1)/n$ times the density of the stream after a time $t_{\rm fill}$ (i.e. when $n=1$ and the stream first began to overlap). Thus we expect similar orbital period variability, except that the amplitude of the variability is expected to be reduced to around $100/n$ per cent. In addition if there were multiple such mass streams, the variability would also be reduced. Thus, for example, the variability seen by \cite{Swan:2019aa}, which corresponds to changes of order 10 -- 30 per cent on timescales of years, could be accommodated in this manner, provided that the last disruption event occurred relatively recently, say within the last few $t_{\rm fill} \sim$ hundreds of years.

To explore this in more detail we perform some simple simulations which capture the expansion of the debris stream over time from the initial disruption to late times ($t\gg t_{\rm fill}$). To do this we initially place a set of particles, which form a spherical, constant density asteroid of radius $R_{\rm ast}$, at the tidal radius (equation~\ref{rt}) with a uniform velocity corresponding to the periastron velocity of the eccentric orbit with $R_{\rm p} = R_{\rm t}$ and $R_{\rm a}=4$\,au. The particles are distributed randomly within the asteroid, following an approximately constant density of $3$\,g/cm$^3$ and a total mass of $10^{20}$\,g. We then integrate the orbits of each particle forwards in time solely under the gravity field of the white dwarf. Thus our simulations correspond to the standard ``frozen-in'' approximation applied to TDEs \citep[e.g.][]{Lacy:1982aa,Lodato:2009aa,Stone:2013aa,Coughlin:2019aa}.  We neglect the self-gravity of the debris and any perturbations from e.g. planets at several au -- we discuss the impact of these approximations in the discussion section. 

\begin{figure}
  \includegraphics[width=\columnwidth]{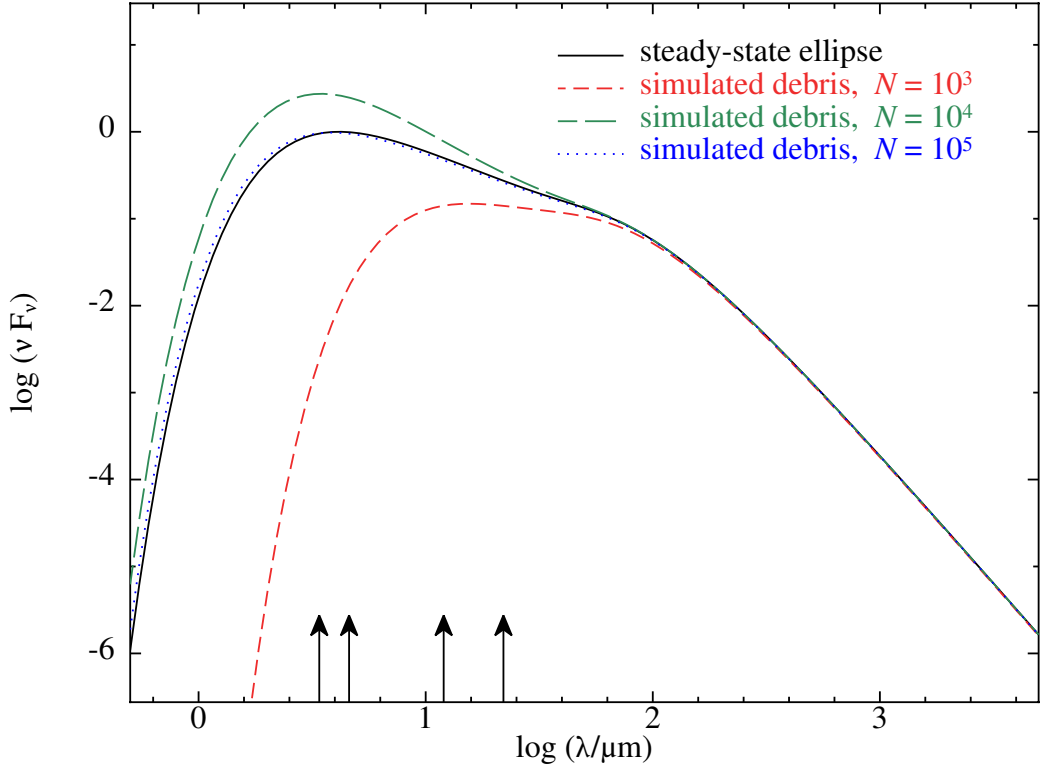}
  \caption{Spectra from the simulated debris streams produced from the tidal disruption of an asteroid by a white dwarf. These spectra are taken towards the end of our simulations, at a time of $4t_{\rm fill}$ ($\approx 93$ orbits) post disruption. At this time the debris should produce a spectrum similar to that of the steady state ellipse from Section~\ref{simplestream}. The fluxes have been scaled to the peak flux for the (fiducial) 4\,au elliptical stream (Fig.~\ref{Fig1}), which is plotted for reference (black solid line). The plotted spectra correspond to asteroids that are modelled with different numbers of particles, with red dashed being $10^3$ particles, green long-dashed $10^4$ particles, and blue dotted $10^5$ particles; in each case the emitting area is kept constant so that any differences are solely attributable to the discreteness of the debris. Note that the $10^6$ particle debris stream spectrum is not shown as it is indistinguishable from the steady-state ellipse spectrum on this plot. The effective wavelengths of the {\it WISE} bands (3.4, 4.6, 12.0, and 22.0\,$\mu$m) are marked with black arrows for reference. This figure shows that the spectrum emitted from the simulated debris stream at late times is similar to that of an ellipse with a constant mass flux of debris around the orbit if sufficient numbers of particles are used. For low numbers of particles ($\lesssim 10^4$) there is substantial variability even at late times, and particularly at short wavelengths, due to the discreteness of the debris passing periastron.}
  \label{Fig2}
\end{figure}

To start with, we plot in Fig.~\ref{Fig2} the spectrum from the simulated asteroid debris at late times ($t = 4t_{\rm fill}$, which is approximately 93 orbits of the original asteroid's centre of mass) for different numbers of particles representing the asteroid (from $10^3$ up to $10^5$ particles\footnote{We have also performed the simulation with $10^6$ particles, but this is not plotted on Fig.~\ref{Fig2} as the resulting spectrum is indistinguishable from that of the $10^5$ particle simulation.}). We also plot on the same figure the spectrum from the corresponding steady-state ellipse from Section~\ref{simplestream} for comparison. The total emitting area in each simulation is the same, and thus differences are solely attributable to the discreteness of the particles representing the stream. For example, the $N=10^3$ simulation happens to have no particle near periastron at the time at which the spectrum is calculated and thus has relatively little flux at short wavelengths. In contrast the $N=10^4$ simulation has several particles near periastron and thus its spectrum has a relatively high amount of flux at short wavelengths. For $N=10^5$ and $N=10^6$ there are sufficient particles around the ellipse that the discreteness is minimal and the spectra are nearly identical to that of the steady state ellipse. Note that small deviations from the steady state ellipse (black solid line) are expected as the simulated debris contains a small spread of orbits and thus a small distribution of temperatures at any radius.

\begin{figure*}
  \includegraphics[width=0.461\textwidth]{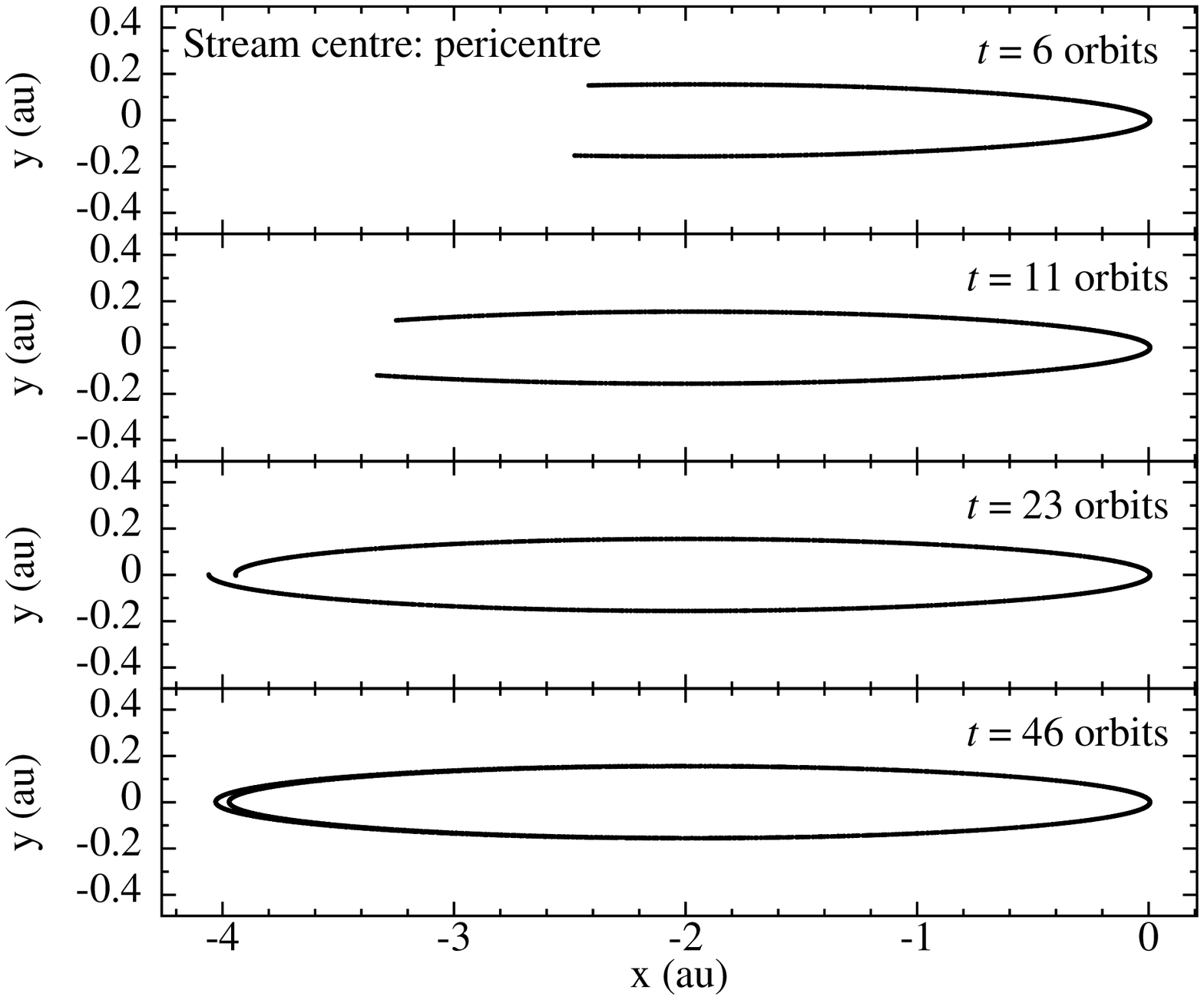}\hfill
  \includegraphics[width=0.515\textwidth]{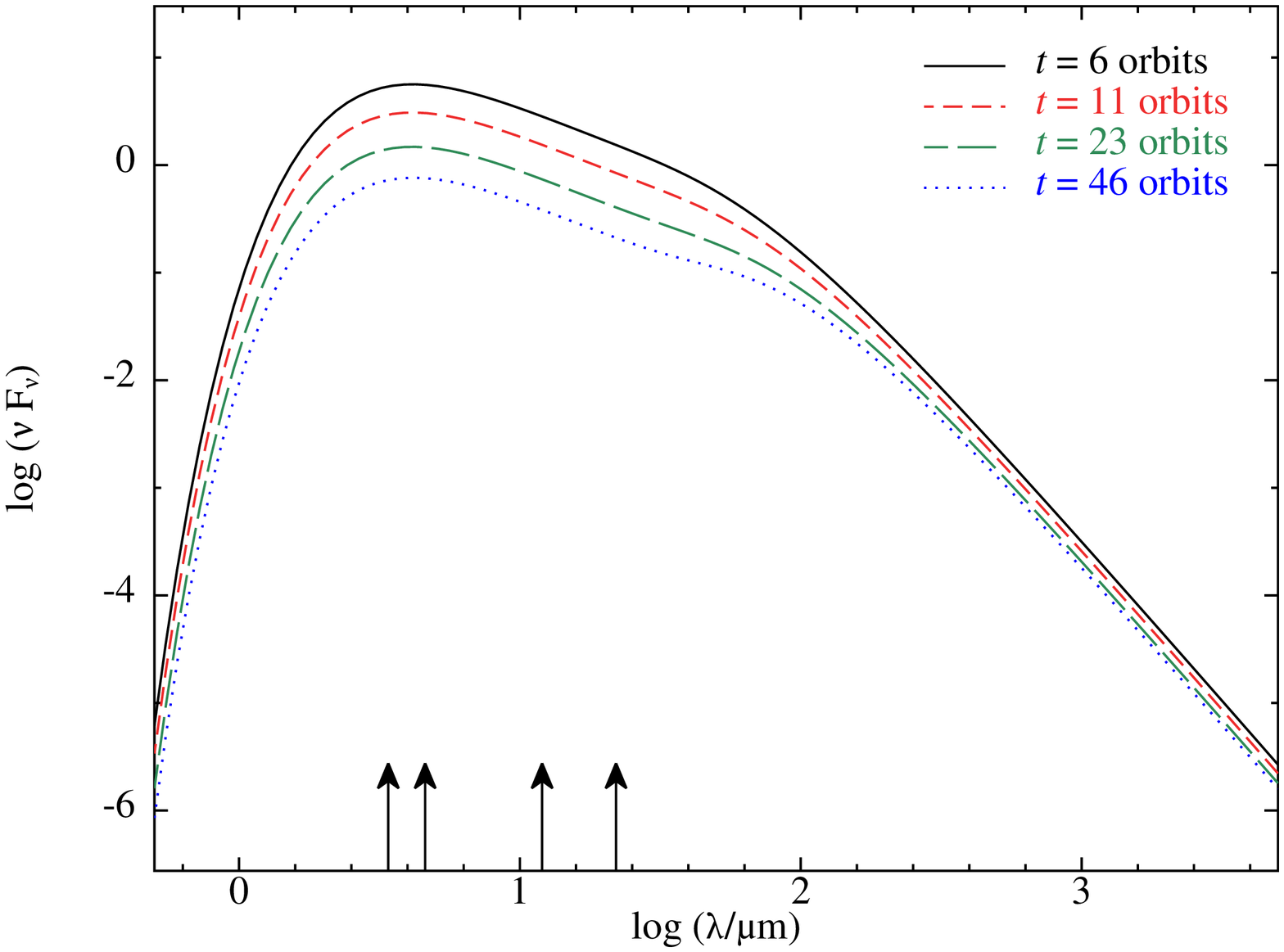}\vspace{0.1in}
  \includegraphics[width=0.461\textwidth]{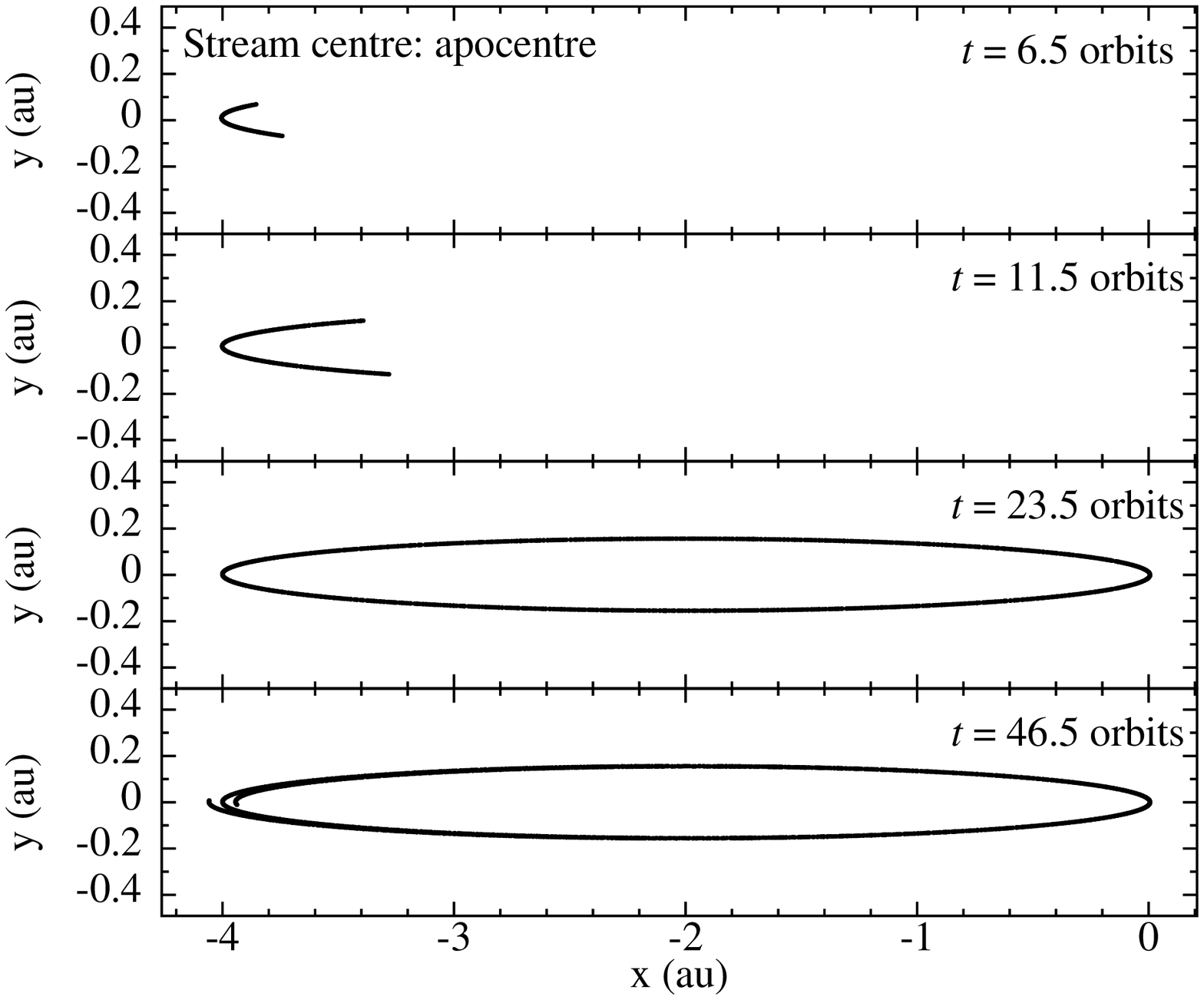}\hfill
  \includegraphics[width=0.515\textwidth]{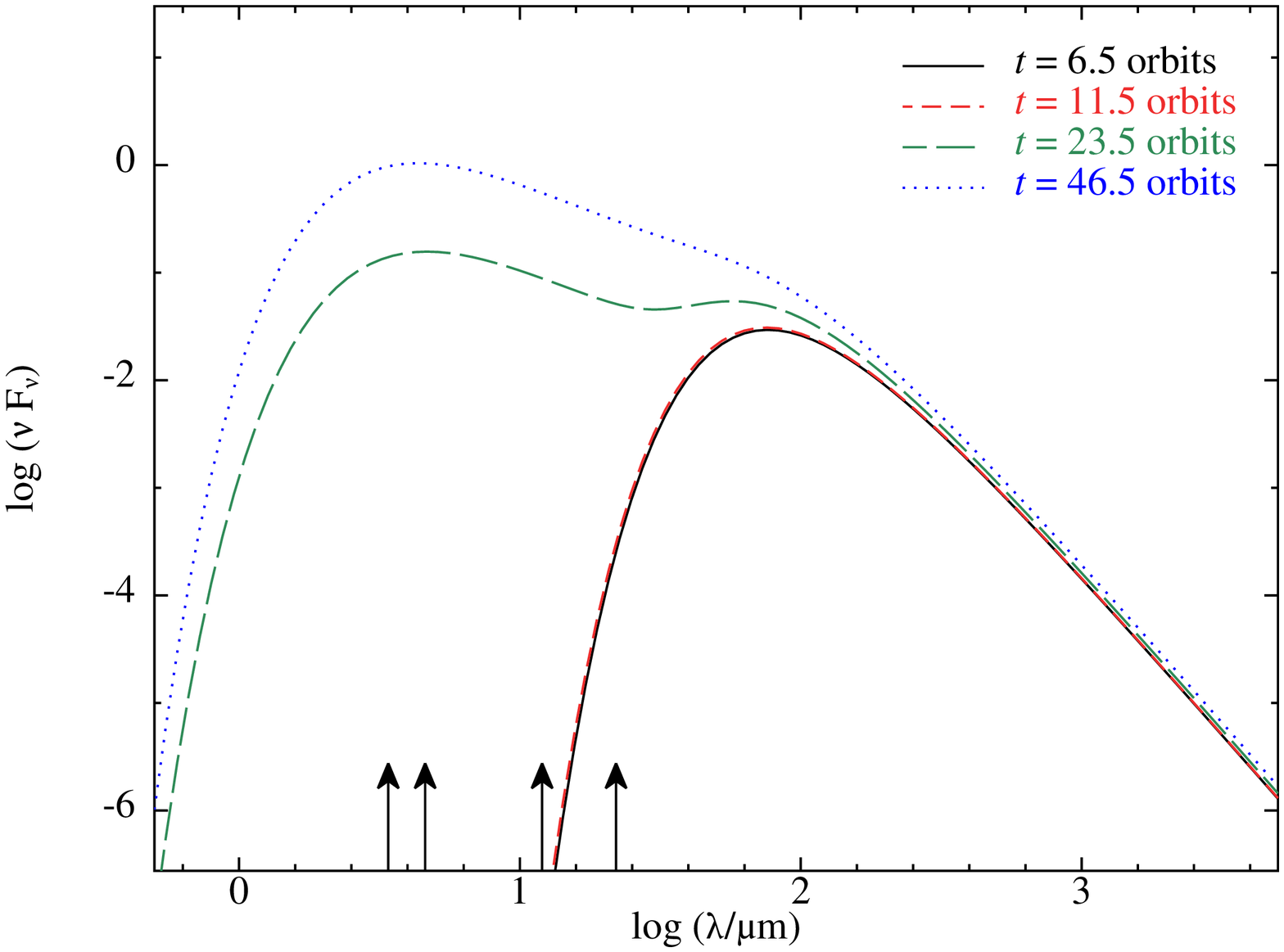}
  \caption{Debris distributions and corresponding spectra for the simulated debris stream with $N = 10^6$ particles. The top left panel shows the debris distribution after 6, 11, 23, and 46 orbits (top to bottom panels), corresponding to timescales of $\approx t_{\rm fill}/4$, $t_{\rm fill}/2$, $t_{\rm fill}$, and $2t_{\rm fill}$. The top right panel shows the spectra resulting from the distributions of debris in the top left panel. These two (top) panels correspond to times where the centre of mass of the original asteroid is at periastron. The bottom left panel shows the debris distribution after 6.5, 11.5, 23.5, and 46.5 orbits (top to bottom panels), corresponding to timescales of $\approx t_{\rm fill}/4$, $t_{\rm fill}/2$, $t_{\rm fill}$, and $2t_{\rm fill}$. The bottom right panel shows the spectra resulting from the distributions of debris in the bottom left panel. These two (bottom) panels correspond to times where the centre of mass of the original asteroid is at apastron. On the right hand panels, the effective wavelengths of the {\it WISE} bands (3.4, 4.6, 12.0, and 22.0\,$\mu$m) are marked with black arrows for reference. The fluxes have been scaled to the peak flux for the (fiducial) 4\,au elliptical stream (Fig.~\ref{Fig1}). The left hand panels show that the debris fills out the ellipse on a timescale of $t_{\rm fill}$ (cf. Section~\ref{simplestream}). The right hand panels show how the spectrum changes as the debris orbits from periastron to apastron (comparison of top to bottom) and over a timescale of several $t_{\rm fill}$.} 
  \label{Fig3}
\end{figure*}

\begin{figure}
  \includegraphics[width=\columnwidth]{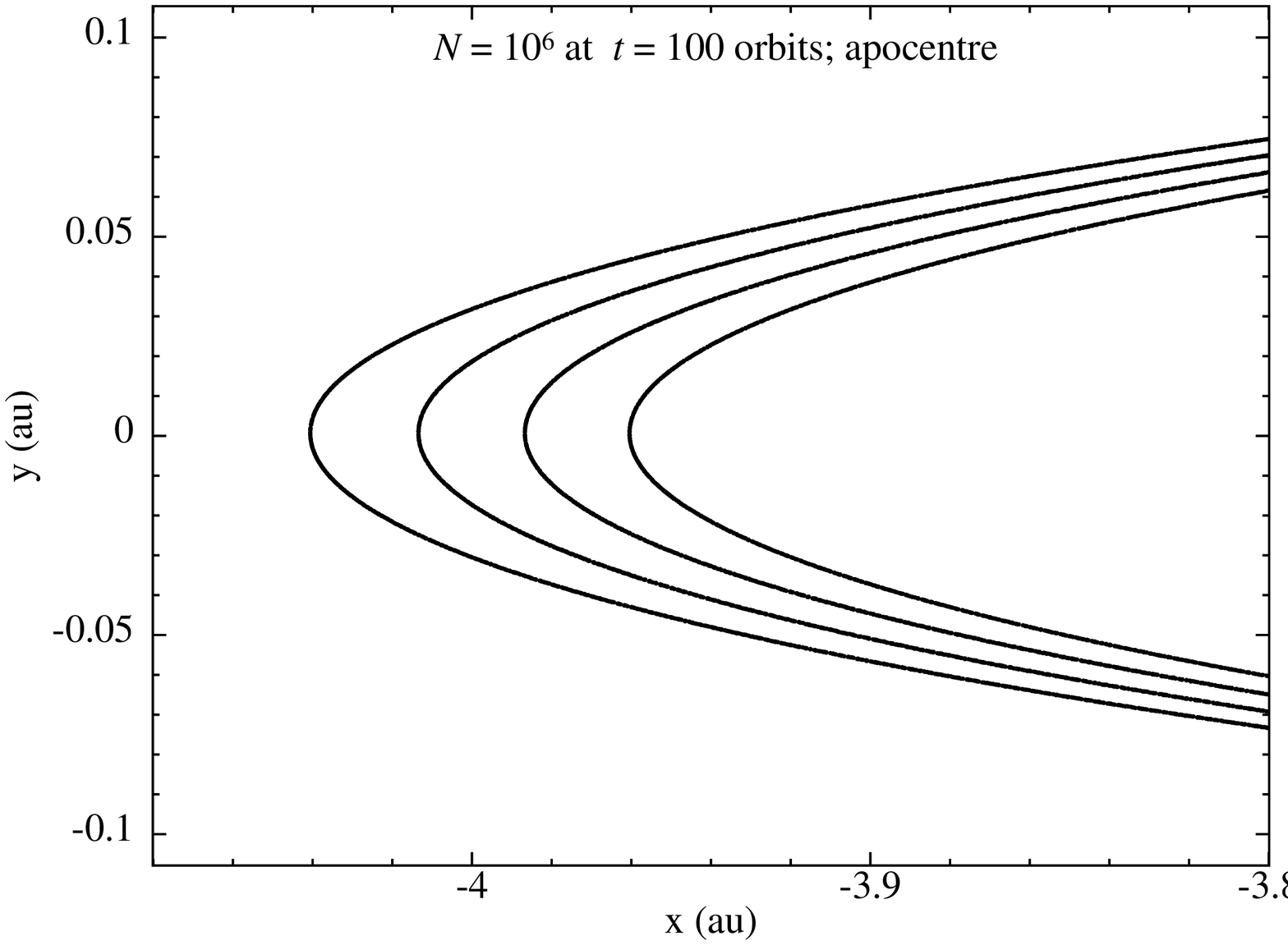}
  \caption{Debris distribution near apastron after 100 orbits, which corresponds to $\approx 4.4 t_{\rm fill}$. The debris traces out a line which wraps around the white dwarf (the white dwarf, not depicted, is located to the right at [0.0,0.0]). The empty space between each wrap is filled by the orbits of other debris which are currently elsewhere on their orbits than  near apastron. Thus each line cycles like a barber pole from the minimum apastron ($r_{\rm a} \approx 3.94$\,au for the leading edge of the stream) to the maximum apastron ($r_{\rm a} \approx 4.06$\,au for the trailing edge of the stream).}
  \label{Fig4}
\end{figure}

In Fig.~\ref{Fig3} we provide debris distributions and spectra when the centre of the original asteroid is at periastron (top panels) and apastron (bottom panels) at several times throughout the simulation corresponding to $\approx t_{\rm fill}/4$, $t_{\rm fill}/2$, $t_{\rm fill}$, and $2t_{\rm fill}$. As expected from the analytical calculations in Section~\ref{simplestream}, the left hand panels show that after a time of approximately $t_{\rm fill}$ the debris has spread out around the ellipse, and that by a time of $2t_{\rm fill}$ the debris has wrapped around twice. (We plot in Fig.~\ref{Fig4} the debris distribution near apastron after 100 orbits, $\approx 4-5t_{\rm fill}$, which shows this effect more clearly.) The corresponding spectra (the right hand panels of Fig.~\ref{Fig3}) show that (a) at periastron the flux at each wavelength slowly decreases over time as more debris is distributed towards apastron where the temperatures are lower and thus the flux is reduced\footnote{Note that a higher temperature black body emits more flux at {\it every} wavelength than a cooler black body of the same emitting area.}, and (b) at apastron the flux at long wavelengths ($\gtrsim 100\,\mu$m) is approximately constant but the flux at short wavelengths increases strongly over time as material spreads out towards periastron where the temperatures are higher. After several $t_{\rm fill}$ the flux at each wavelength becomes approximately constant (assuming sufficient numbers of particles are used to represent the stream), which occurs when the debris has reached an approximate steady state in terms of mass flux around the ellipse (cf. equation~\ref{masstheta}).

\begin{figure*}
  \centering   
  \begin{overpic}[width=\textwidth]{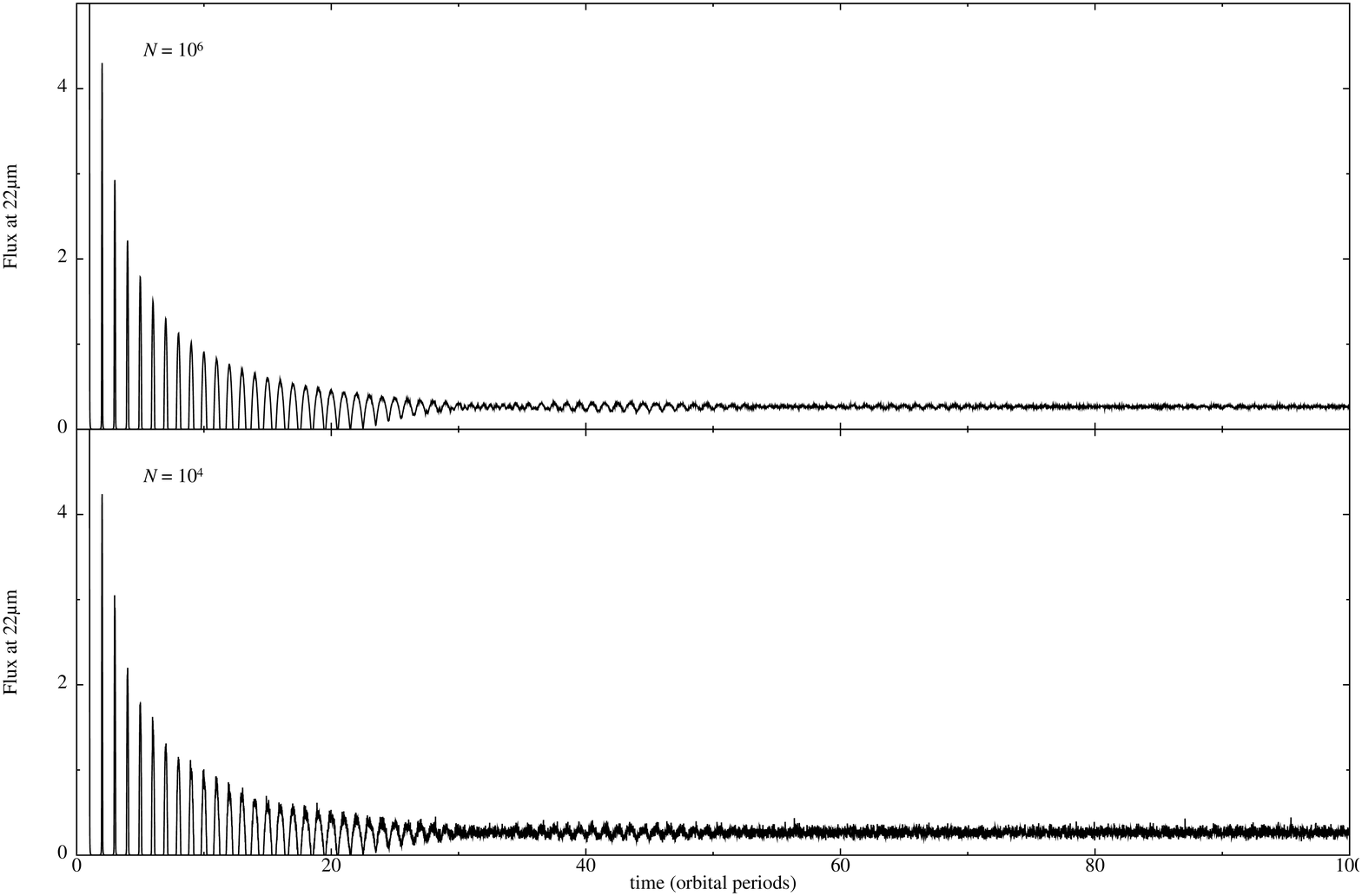}
     \put(45,40.5){\includegraphics[width=0.5\textwidth]{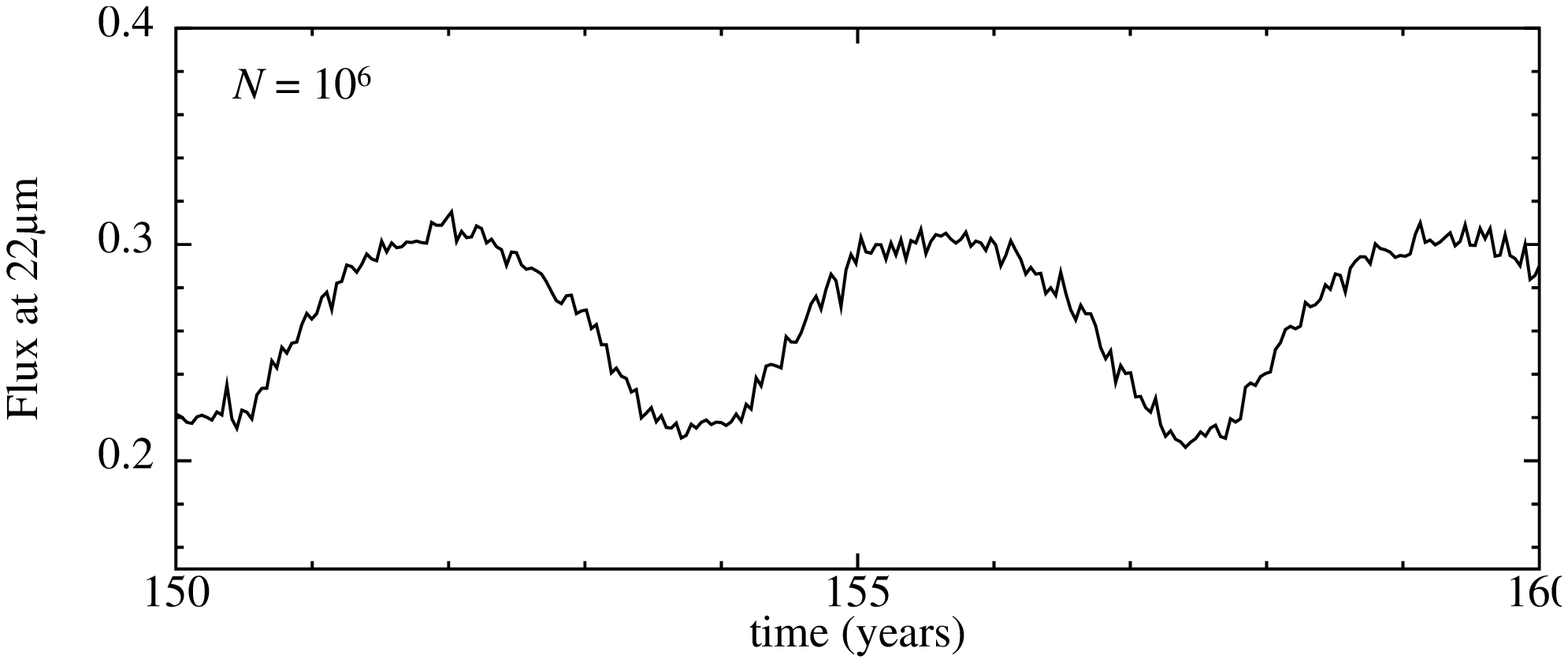}}  
     \put(45,9){\includegraphics[width=0.5\textwidth]{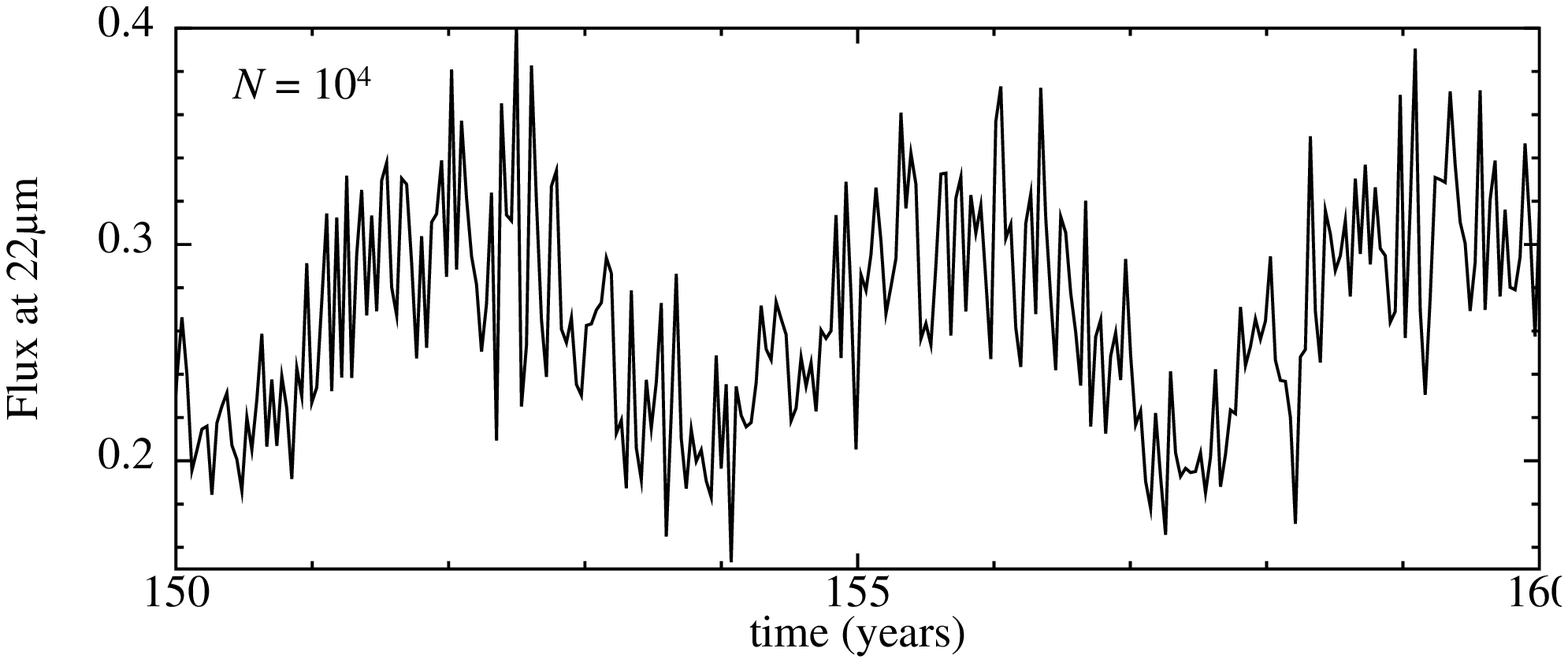}}  
  \end{overpic}
  \caption{The flux at 22\,$\mu$m from the simulated debris as a function of time since disruption, where the debris is modelled with $10^6$ particles (top panel) and $10^4$ particles (bottom panel). The insets in each panel show a zoom-in on the lightcurve between a time of 150 years post-disruption and 160 years. The fluxes have been scaled to the peak flux for the (fiducial) 4\,au elliptical stream (Fig.~\ref{Fig1}). The time axis on the main plots is given in orbital periods of the original centre of mass of the disrupted asteroid, while in the insets the time is given in years. The conversion factor between orbital periods and years is given by equation~\ref{P}, with one orbit corresponding to $\approx 3.66$ years for the parameters of the simulation. The lightcurve in the top panel, corresponding to $10^6$ particles, shows large amplitude, coherent variability (of order 100 per cent) on the orbital period for times $\lesssim t_{\rm fill} \approx 23$ orbits. After this time, the lightcurve variability is significantly reduced, and at late times becomes approximately constant. The lightcurve displays some noise due to discrete particles passing periastron---where both the particle velocity and temperature are highest---and the amplitude of the (Poisson) noise decreases with increasing particle number. It is also possible to see an additional reduction in the amplitude of the variability between 30 and 35 orbits and between 55 and 60 orbits, caused by a beat between the orbital period and the filling timescale (see text for discussion).}
  \label{Fig5}
\end{figure*}

In Fig.~\ref{Fig5} we plot the lightcurve from our simulation in the {\it WISE} 4 band. For simplicity, we create the lightcurve using the value of the flux at the band's effective wavelength of $22$\,$\mu$m, rather than employing the relevant {\it WISE} response function over the full band -- given our simplistic approach to treat the emission as a sum of black bodies such an extra complication is not warranted. The lightcurve in the top panel of Fig.~\ref{Fig5} corresponds to $10^6$ particles, while the lightcurve in the bottom panel corresponds to $10^4$ particles. These lightcurves show large amplitude, coherent variability (of order 100 per cent) on the orbital period for times $\lesssim t_{\rm fill} \approx 23$ orbits. After this time, the lightcurve variability is significantly reduced, and at late times the lightcurve becomes approximately flat at a non-zero flux value. The two insets show a zoom-in of the time between 150 years ($\approx 41$ orbits) and 160 years ($\approx 44$ orbits) post-disruption. These show that continuous observation of the stream is required for several years to determine the lightcurve shape, and that for significant periods of time (of order 1-2 years) the lightcurve may appear to be linearly increasing or linearly decreasing with time.

The lightcurve in Fig.~\ref{Fig5} displays some noise due to discrete particles passing periastron---where both the particle velocity and temperature are highest---and the amplitude of the (Poisson) noise decreases with increasing particle number. This increase in the amplitude of the noise at smaller particle numbers indicates that the degree to which the original asteroid is disrupted is imprinted on the lightcurve; if the asteroid is obliterated into many tiny particles (say $\gg 10^6$) then the resulting lightcurve is relatively free of noise. However, if the asteroid breaks up into a modest (say $\lesssim 10^3$) number of chunks, then the resulting lightcurve will become dominated by the discreteness of these chunks passing periastron on a timescale of order $t_{\rm fill}$. Each of these `events', where a discrete chunk sweeps through periastron, will be accompanied by a burst of additional flux at short wavelengths with a characteristic timescale of order a few times $(R_{\rm p}^3/GM_{\rm wd})^{1/2}\approx 1$\,hr for $M_{\rm wd}=0.6 M_\odot$ and $R_{\rm p} = R_{\rm t}$, but not necessarily with any associated periodicity. Thus it is possible that particularly high-cadence lightcurves may be sensitive to this type of variability, which would be lost in coarser data with bin widths of several hours where it would appear as noise.

There is an identifiable beat between the orbital period and the filling timescale in the lightcurves in Fig.~\ref{Fig5}: the lightcurve is essentially flat (dominated by noise) between 30 and 35 orbits and also between 55 and 60 orbits, while after these sections the signal (periodic on the orbital period) returns (e.g. from 35 to 50 orbital periods) with decreasing amplitude on each return. This beat causes a switch in the location of the peak flux from being on integer multiples of the orbital period to half integers following a period of two peaks per orbit. This can be understood as follows: In the simulated debris stream the mass per unit length is not constant along the stream. If one defines the distance from the centre of mass in the original asteroid in the direction pointing towards the white dwarf with the variable $\mathcal{R} = \xi R_{\rm ast}$ where $\xi \in [-1,1]$, then the mass of the (constant density) asteroid located at different $\xi$ values forms a dome with its peak at $\xi=0$ and goes to zero at $\xi=\pm1$. Therefore the highest density on the stream occurs at $\xi=0$, i.e. coincident with the centre of mass of the original asteroid. For times $<t_{\rm fill}$, this overdensity is responsible for the peaks in the lightcurve that line up with integer multiples of the orbital period, i.e. every time the overdensity returns to periastron. After approx $t_{\rm fill}$ the two ends of the debris stream begin to overlap, and thus return to periastron at approximately the same time causing a secondary peak of initially small amplitude. This secondary peak occurs at a phase of $\approx \phi_{\rm com} + \pi$, where $\phi_{\rm com}$ is the phase of the primary peak caused by the centre of mass of the original asteroid -- thus the secondary peak lags the primary peak by $\approx t_{\rm orb}/2$. As the ends overlap further the amount of debris starting to overlap increases, i.e. the $|\xi|$ value at the point of overlap decreases, and the secondary peak therefore becomes larger than the primary peak. By a time of approximately $1.5t_{\rm fill}$ the overlapping region of the stream advances to overlap with the original centre of mass of the stream, thus returning the peak flux to integer values of the orbital period. We note that, while such a commensurability is found by our idealised simulations, it is likely that such a signal would be lost in a realistic system where the simple orbits proposed here are affected by other gravitating bodies (see the discussion).

So far we have only discussed the lightcurve plotted in Fig.~\ref{Fig5} which is at a wavelength of $22$\,$\mu$m corresponding to the {\it WISE} 4 band. As several other wavelengths are of interest for studying the infrared excess observed in white dwarf spectral energy distributions, we consider these here. These include the near-infrared K band (2.2\,$\mu$m), the mid-infrared {\it WISE} bands at 3.4\,$\mu$m, 4.6\,$\mu$m and 12\,$\mu$m\footnote{There is typically a strong peak of emission observed in this band attributed to silicate emission. Our simple estimates of the spectra do not take this into account.}, and the far-infrared at wavelengths $\sim$100\,$\mu$m\footnote{E.g. for data from ALMA and future instruments such as SPICA. We note that some observations of dusty white dwarfs have been performed at far-infrared wavelengths, e.g. for G29-38 \citep[e.g.][]{Farihi:2014aa} and in general the fluxes are all upper limits.}. The lightcurves in these bands (not shown) all display the same time dependence as the lightcurve at $22$\,$\mu$m shown in Fig.~\ref{Fig5} -- that is to say that the peaks and troughs of the lightcurve variability all occur in every band at the same time, and this is driven by the increase in flux as material orbits past periastron. A higher fraction of the flux at shorter wavelengths arises from the emission at periastron, and thus the shorter wavelength lightcurves display larger amplitude variability at all times (from the periodic signal and from the noise), and they have a shorter time at which the lightcurve begins to be dominated by the noise from discreteness at periastron. The orbit averaged flux in each band is essentially a constant -- it would be exactly a constant if the debris orbits were exactly the same (i.e. all the debris had the same semi-major axis and eccentricity, and thus orbital period) as the emission from each piece of debris over the whole orbit is the same, independent of where on the orbit it started. However, because the orbits are slightly different the flux averaged over the orbital period of the centre of the original asteroid is approximately constant with a small degree of scatter. For the simulations presented here, the fluxes in the 2.2\,$\mu$m, 4.6\,$\mu$m, 22\,$\mu$m and 100\,$\mu$m bands are in the approximate ratios of 1\,:\,1.80\,:\,0.47\,:\,0.10 respectively.

\section{Discussion}
\label{discussion}
In this paper we have explored a model for the tidal disruption of an asteroid  by a white dwarf. Typically we expect these asteroids to originate from radii of order a few au \citep{Debes:2002aa,Frewen:2014aa}. To be disrupted by the white dwarf such planetary material must reach the tidal radius, of order $R_\odot\approx 100R_{\rm wd}$ (see equation~\ref{rt}). Upon reaching this radius, the asteroid is expected to break up into a debris stream \citep{Jura:2003aa,Debes:2012aa,Veras:2014aa}.

Following the results of the simulations of the tidal disruption of rubble pile asteroids by \citet[][see also \citealt{Veras:2014aa}]{Debes:2012aa} we assumed that the asteroid is broken into discrete pieces that move solely under the gravity of the white dwarf, and thus over time they spread to fill their orbital ellipse. Under this assumption\footnote{We note that there is not genuine consensus in the literature that this will occur in every scenario \citep[see e.g.][]{Hahn:1998aa}. We discuss this further in Section~\ref{caveats}.} we predicted, and verified with simple numerical simulations, the timescale and properties of the debris from the point of disruption until late times when the debris has approached an approximate steady state in terms of mass flow around the orbits. Because the energy spread of the debris is much smaller in magnitude than the orbital energy of the original asteroid, the filling timescale is several tens of orbits and the ensemble of debris occupy very similar orbits (their semi-major axes and eccentricities varying by $\sim 1$ per cent and $\sim 0.1$ per cent respectively). Assuming a basic model for the absorption of flux from the white dwarf by the debris and the re-emission of this energy as a simple black body, we explored the spectral and time variability of the debris. At each wavelength the flux is highly variable (up to 100 per cent) for times $\lesssim$ several $t_{\rm fill}$. At later times the lightcurves are essentially flat and become dominated by discreteness effects as individual parts of the asteroid pass by periastron where the temperatures, and thus fluxes, are highest. Our predicted spectra are similar to those observed for the infrared excess in polluted white dwarfs; the flux becomes noticeable at wavelengths $\gtrsim 2$\,$\mu$m, the peak flux is around a few $\mu$m and there is significantly less flux at wavelengths $\gtrsim 100$\,$\mu$m. We do, however, find that the ratio of the maximum to minimum temperature of the debris, $T_{\rm max}/T_{\rm min}$, is typically $\approx 10-40$ for asteroids originating from $\approx 1-10$\,au. This is at least a factor of a few higher than is used in existing models to fit the observed infrared excesses, which typically have $T_{\rm max}/T_{\rm min} \approx 2$ \citep[see e.g.][]{Farihi:2016aa}. The quality of the data at longer wavelengths ($\gtrsim 10\,\mu$m) is poorer than that at shorter wavelengths, however, and there is not yet a strong constraint on $T_{\rm max}/T_{\rm min}$ from the available data.

The main thrust of our paper is to propose that the observations, both spectral and temporal, of the infrared excesses observed in polluted white dwarf spectral energy distributions may be adequately explained by a model in which the asteroid debris/dust remains on the highly elliptical orbits with which it is born. The general consensus in the field is that asteroids/planets etc that were located within a few au would be evacuated during the red giant phase, thus the disrupted asteroids must begin with an orbit originating from beyond several au. The asteroids can be scattered in to the tidal radius around the white dwarf \citep{Debes:2002aa,Frewen:2014aa}, but there is no satisfactory mechanism by which the large amount of orbital energy left in the debris can be efficiently removed to circularise the debris into a disc at a radius of order the tidal radius. Thus the simplest possibility may be that, at least until stream-stream interactions become important, the debris remains in highly elliptical orbits. This may account for the, albeit limited, variability observed to date in these systems, which would not be straightforward to produce in a disc model where the debris executes slowly changing circular orbits. Some authors have attempted to explain the formation of circular (or near-circular) dust discs \citep[e.g.][]{Veras:2017aa,Duvvuri:2020aa} by placing an asteroid on a (near) circular orbit at or within the tidal radius. The physical nature of the initial conditions of such models has yet to be established.

In the following subsections we provide some discussion of effects that may enhance or significantly change the picture we have described in this paper. After this, we draw our conclusions in Section~\ref{conclusions}.

\subsection{Multiple asteroids}
\label{mult_ast}
We have seen above that the timescale for producing an ellipse of debris from a disrupted asteroid is of order one hundred years (equation~\ref{tfill}). The timescale on which these elliptical streams evolve may be much longer, particularly in the absence of gravitational perturbations from e.g. planets at several au (we discuss this possibility below).

For example relativistic apsidal precession of the eccentric orbits due to the gravity of the white dwarf occurs at a rate \citep{Einstein:1915aa}
\begin{equation}
  \Omega_{\rm ap,GR} = \frac{3GM_{\rm wd}}{ac^2(1-e^2)}\sqrt{\frac{GM_{\rm wd}}{a^3}}\,,
\end{equation}
which leads to a timescale of
\begin{equation}
  \label{grprec}
  t_{\rm ap,GR} = \frac{2\pi}{\Omega_{\rm ap,GR}} = 2.5\,{\rm Myr}\,\left(\frac{a}{2\,{\rm au}}\right)^{5/2}\left(\frac{M_{\rm wd}}{0.6M_\odot}\right)^{-3/2}\left(\frac{1-e^2}{1-0.997^2}\right)\,,
\end{equation}
which is far longer than e.g. $t_{\rm fill}$.

Similarly the orbits of the debris may evolve due to Poynting-Robertson (PR) drag from the white dwarf radiation field. Scaling the results from \cite{Wyatt:1950aa} we have that the time to circularise the debris (to approximately its original periastron) is
\begin{equation}
  \label{tcirc}
t_{\rm circ,PR} = (1.7-26.3)\,{\rm Myr}\,\left(\frac{L}{L_{\rm wd}}\right)^{-1}\left(\frac{s}{{\rm cm}}\right)\left(\frac{\rho}{3\,{\rm g/cm}^3}\right)\left(\frac{a_0}{2\,{\rm au}}\right)^{2}\,
\end{equation}
where we have assumed a debris eccentricity\footnote{\cite{Wyatt:1950aa} only list values for an integral, denoted $G(e_0)$, up to $e_0=0.99$. Here we have $G(0.997)=12.2$.} of $e=0.997$, scaled our results to cm sized pebbles of density $3$\,g/cm$^3$, and the range of timescales comes from taking a range of white dwarf temperatures of $12,500-25,000$\,K leading to a range of luminosities of $L_{\rm wd} \approx 4\pi R_{\rm wd}^2\sigma T_{\rm wd}^4 \approx 8.5\times 10^{30}-1.4\times 10^{32}$\,erg/s. Again, these timescales are too long to affect the stream on timescales $\sim t_{\rm fill}$. We do have some freedom of parameters in equation~\ref{tcirc}. The semi-major axis could be significantly larger than $2$\,au, but this would increase $t_{\rm circ,PR}$. The density could be, say, between $0.5-5$\,g/cm$^3$ based on solar system asteroids/rocky objects but this is only a factor of a few. We could significantly shorten the PR drag timescale by assuming the debris could be ground down into micron sized dust (which may itself require long timescales for e.g. collisions to sufficiently fragment the debris) which would reduce the timescale by a factor of $\sim 10^4$, and thus $t_{\rm circ,PR}$ becomes several to tens of $t_{\rm fill}$. This may provide a promising mechanism for circularisation and accretion of dust, but requires that a substantial fraction of the mass of the asteroid be turned into dust within tens of orbits. This has yet to be shown.

So it appears plausible that, if the debris stream can avoid significant perturbation from the larger (few au) scale planetary system, then elliptical debris streams may persist for many $t_{\rm fill}$ and possibly for longer than the recurrence time, $t_{\rm recur}$, for another asteroid to be kicked in and disrupted.

We can estimate the recurrence time implied by the observed accretion rates of metals on to the white dwarfs, which are ${\dot M}\sim 10^6-10^{10}$\,g/s. Assuming that these rates are approximately constant in time implies that an asteroid of mass, say, $10^{20}$\,g must be accreted every $t_{\rm recur} \sim (10^3-10^7)(M_{\rm ast}/10^{20}\,{\rm g})$\,years. If the accretion rates are caused by the cumulative effect of $n$ debris streams, rather than a single stream, then this becomes $(10^3-10^7)(M_{\rm ast}/10^{20}\,{\rm g})/n$\,years. Thus in the systems with the highest accretion rates it is plausible that the disruptions are occurring on a timescale $\lesssim t_{\rm fill}$, while for the systems with the lowest accretion rates, we expect the streams to reach a steady state before the next asteroid is delivered. These two cases exhibit distinct variability properties. In both cases, in general, there will be a level of background flux from `old' streams that can dampen the orbital timescale variability from a `new' stream. If $t_{\rm recur} < t_{\rm fill}$ we would expect to see continuous variability at similar timescales (the orbital periods of each asteroid will not vary much, assuming they come from the same belt and are caused by the same perturber) but that appears stochastic rather than periodic as the flux peaks from the periastron passage of the bulk of the material in the debris streams will be uncorrelated. For $t_{\rm recur} > t_{\rm fill}$ we would expect to see, across a sample of objects, a fraction of order $t_{\rm fill}/t_{\rm recur}$ of objects displaying orbital period variability and the rest displaying approximately constant flux in the infrared bands. For a more in depth discussion of these (and broader) considerations relating to the recurrence time, see \cite{Wyatt:2014aa}.

If the number of coexisting streams grows to large values then we can expect that collisions between the streams will occur. Such collisions occur predominantly at periastron where the densities are highest and with relative velocities of the colliding debris that are of order the orbital velocity of the debris. Such collisions will instantly vaporise the debris, and thus provide a source of gas near periastron. We discuss gas in these systems next.

\subsection{Implications for observations of gas discs}
In a small number of polluted white dwarfs there is clear observational evidence for the presence of gas at radii of order the tidal and/or sublimation radius. \cite{Gansicke:2006aa} reported the first such observation, where they discovered a Ca~II emission triplet in the spectrum from the hot, young white dwarf J1228+1040 (with $M_{\rm wd} = 0.77M_\odot$, $R_{\rm wd} = 0.011R_\odot$ and $T_{\rm wd} = 22,000$\,K) implying the presence of a metal-rich gas disc. The line profiles are as expected for a Keplerian disc, but with a significant asymmetry that suggests the disc is eccentric. Modelling of this disc yields orbital radii of $\approx 0.64R_\odot-1.2R_\odot$ and an eccentricity of $0.21$. Similar properties are found for J0845+2257 and J1043+0855 \citep{Gansicke:2008aa,Gansicke:2008ab}. Subsequent observations of these three systems were performed by \cite{Melis:2010aa}. \cite{Farihi:2012ab} report the discovery of a fourth system displaying the (again asymmetric) Ca II triplet, this time around a cooler white dwarf of about $13,300$\,K and they suggest that this implies that the gas in dust discs is therefore not related to (at least in all cases) sublimation of the dust. \cite{Brinkworth:2012aa} present {\it Spitzer} data for four white dwarfs known to host gas discs. More recently it has become possible to look at the time evolution of the Ca II triplet, finding that the lines evolve in a way that is commensurate with precession of the eccentric gas disc on timescales of tens of years \citep{Manser:2016aa,Cauley:2018aa,Dennihy:2018aa,Manser:2019aa}. The first statistical study of the occurrence rate of gaseous components in dusty white dwarfs is provided by \cite{Manser:2020aa}, who estimate that $4^{+4}_{-2}$ per cent show evidence for gas.

The standard model for pollution of white dwarf atmospheres assumes that the debris formed from the tidal disruption of an asteroid ends up orbiting in a (near) circular disc at a radius of order the tidal radius \citep{Jura:2003aa}. For such a dust disc accreting inwards on to the white dwarf \citep[cf.][]{Rafikov:2011aa,Rafikov:2011ab} it is inescapable that gas will be produced at the sublimation radius (always greater than the white dwarf radius). However, we only see the gas in a small fraction of objects. While in some systems the gas may just be difficult to detect, the current low observed occurrence rate of gas discs poses a problem for this standard dust disc model. Further, the gas produced at the sublimation radius, from a circular dust disc, occupies circular orbits. Therefore the gas distribution is expected to be approximately time-steady, varying only on the drain timescale of the dust disc. Again, this does not appear consistent with the observed eccentric and precessing gas discs.

If the asteroid debris remains on highly eccentric orbits, as proposed here, gas may be produced in several ways. The debris can be vaporised by collisions between the debris occurring at sufficiently different orbital phases. This requires the debris to be scattered from a simple ellipse, potentially by the remaining planetary system at several au (including the perturber that drove the asteroid in in the first place), or that multiple debris streams exist and that two or more of their orbits overlap somewhere. Alternatively, the periastron of the asteroid debris may lie inside the sublimation radius, $R_{\rm sub}$\footnote{Note that as the radius of sublimation depends on both grain size and composition, there is not a single specific location but more likely a broad radial range where sublimation can occur. For our discussion above, $R_{\rm sub}$ may be thought of as the largest radius inside which a reasonable amount of sublimation would occur.}, allowing a fraction of the debris to be turned into gas during every periastron passage. If we have $R_{\rm p,ast} \sim R_{\rm t} > R_{\rm sub}$ then to produce gas we require additional perturbations to the debris stream orbits (as in (1) above) to further reduce the periastron of the debris orbits until $R_{\rm p}\lesssim R_{\rm sub}$. Some orbits will directly impact the white dwarf, but others will have $R_{\rm wd} < R_{\rm p} < R_{\rm sub}$. Debris on these orbits will lose material to gas every periastron passage. If, instead, we have $R_{\rm wd} < R_{\rm p,ast} < R_{\rm sub}$ then, as the asteroid is originally disrupted into debris, significant amounts of gas will also be produced. Any remaining rocky debris will be further degraded into gas over subsequent periastron passages. The presence of gas near periastron may also help to bring in additional dust through drag with the gas disc. It would appear likely that the production of significant amounts of gas requires the periastron of the asteroid orbit to lie within the sublimation radius, defined by $\beta_{\rm c} = R_{\rm t}/R_{\rm sub}$. This may account for the small number of systems that show strong gas emission, as disruptions with such small periastra (large $\beta$) are unlikely. Each of these mechanisms lead to the production of gas in a manner that is both stochastic and time variable. Further, the increase in the sublimation radius for hotter white dwarfs suggests that the presence of gas should be biased towards these systems. We also note that once the elliptical debris is in the form of gas (also, initially, on highly eccentric orbits), the scale height of the gas (stream width) is larger due to gas pressure; this makes subsequent collisions of the gas streams (shocks) more likely. Shocks in the gas flow lead to orbital energy loss and thus circularisation over time. Thus the observed gas would be expected to have moderate, time-dependent eccentricities.

In summary, dust discs are always expected to produce gas at $R_{\rm sub} > R_{\rm wd}$, and will do so on circular orbits and at an approximately constant rate. This appears at odds with the observed eccentric, variable gas. In contrast, the elliptical debris model presented here is expected to produce gas intermittently on eccentric orbits that subsequently circularise over time. This, at least at first glance, appears commensurate with what is observed.

\subsection{Model simplifications}
\label{caveats}
Throughout this work we have made a number of simplifications. These can be broadly split into three categories, (1) treatment of the debris, (2) treatment of the system, (3) treatment of the spectra, and we discuss each of these in turn.

Our simple treatment of the debris ignores several possible effects. We have treated the debris as test particles orbiting in the gravity of the white dwarf. This is motivated by the assumption that other relevant forces (such as self-gravity) are overcome by the tidal field of the white dwarf during the disruption process, and appears to be consistent with the calculations of e.g. \cite{Veras:2014aa}. In reality the debris can undergo collisions, which may lead to fragmentation into smaller objects and exchange of energy and angular momentum within the stream. While the stream follows approximately the same orbits collisions should be weak, but if the orbits are perturbed by, say, other gravitating bodies then collisions may become important. We have neglected the self-gravity of the debris, mainly for computational simplicity. We anticipate that self-gravity will change the detail of this picture as it is possible that the debris stream could fragment into bound pieces on the return to apastron \citep[see e.g.][]{Hahn:1998aa}, which is also known to occur in the stellar TDE case \citep{Coughlin:2015aa,Coughlin:2020aa}. We will explore the effects of self-gravity in future work. We have treated the debris as single-sized, when in reality the disruption of the asteroid will yield debris of varying sizes. Our spectra and lightcurves are all scaled to a given emitting area for the debris, so in this regard they can be thought of as representing whichever debris size (taking into account the mass-size distribution) yields the largest emitting area -- this is usually the smallest debris. Finally we have assumed that the asteroid is spherical, constant density, homogeneous and non-rotating. Each of these properties is violated by realistic asteroids, but we anticipate including such effects will not change the basic picture outlined here. For example, including the rotation of the disrupting object can broaden the resulting energy distribution by a modest amount that is likely to be degenerate with, e.g., disrupting a larger sized object \citep[cf.][]{Golightly:2019aa}.

Our simple treatment of the white dwarf system is perhaps the most worrisome simplification to the problem. In this work, similar to many other works in the literature, we have treated the asteroid and white dwarf in isolation. In reality the white dwarf is host to other gravitating bodies that can affect the debris orbits. For example, the asteroid belt from which the asteroid likely originated must occupy orbits at the apastron of the debris ellipse. Similarly the planetary perturber that knocked the asteroid on to an orbit which passed close enough to the white dwarf to be disrupted must exert a strong gravitational influence in this region -- once the debris fills its ellipse the perturber will deflect the material at apastron once per orbit of the perturber (and such gravitational interactions with the perturber may also occur before the debris fills the ellipse). This may serve to isotropise the orbits of the debris, such that they more resemble a cloud of debris, increasing the chance for collisions between the debris streams. Further the cumulative effect of such perturbations may decrease the periastron of a significant amount of material leading to direct impact of debris with the white dwarf surface. It is possible that if enough debris were orbiting the white dwarf, then this mechanism could become a dominant mode of delivery of metals to the white dwarf surface (or close enough to become trapped in the white dwarf magnetosphere; \citealt{Farihi:2017aa}). We have simulated one value of the periastron distance for the asteroid. For $R_{\rm wd} \approx 0.01R_\odot$ and $R_{\rm t} = 1.3R_\odot$, the periastron distance can yield $\beta \equiv R_{\rm t}/R_{\rm p} = 1-130$. The energy spread for different $\beta$ values is expected to be approximately the same (see Section~\ref{simplestream}) and thus the picture we have presented remains the same. However, smaller periastron orbits may allow for direct sublimation at periastron, increased rate of collisions grinding debris into dust, and shorter PR drag timescales. We have assumed that relativistic precession of the orbits from the white dwarf's gravity is unimportant (cf. equation~\ref{grprec}), which is a good approximation unless the stream has a very small periastron (for $R_{\rm p}\sim R_{\rm wd}$, equation~\ref{grprec} yields $\approx 2\times 10^4$\,yrs) or has very long lifetimes $\left(\sim{\rm Myr}\right)$.

Finally, we have employed a simple model for the emission from the debris, i.e. that of a black body spectrum. In reality dust spectra are complex, and depend strongly on the composition of the debris. The main place this will make an order unity difference is in predicting the flux at $\approx 8-12$\,$\mu$m (e.g. the {\it WISE} 3 band) where there is a strong silicate feature. For the majority of the currently available data, particularly the photometry, there is insufficient sampling at different wavelengths to warrant a significantly more complex approach in this first investigation. To calculate the spectra and lightcurves we have also assumed that the debris is optically thin and visible to the observer at all times. Soon after disruption the debris will be optically thick, and will become optically thin once it has spread out to a degree around the orbit. If sufficiently many debris streams coexist from multiple asteroid disruptions then they may begin to block out a significant amount of the flux from the white dwarf, which should be readily apparent at shorter wavelengths. It is also possible for the debris to transit the white dwarf, leading to additional sources of variability; assuming that the debris maintains a vertical distribution of order the original asteroid size and is comprised of mostly small particles so that we might hope for optical depths approaching unity, then once it is spread around its orbit it will block a fraction of the white dwarf flux of order $R_{\rm ast}/R_{\rm wd} \approx 0.3 (R_{\rm ast}/20\,{\rm km})$ per cent. Thus a large asteroid, or additional stirring of the debris orbits, would be required for a significant transit. A transit of the type discussed here may have been observed by \cite{Vanderbosch:2019aa}.

\section{Conclusions}
\label{conclusions}
It is well-known that a substantial fraction of white dwarfs, up to 50 per cent, are polluted in the sense that they show atmospheric absorption lines of material with abundances matching that found in asteroids/rocky planets. This finding indicates ongoing accretion of such material. Several per cent of white dwarfs show excess infrared flux from several to several tens of microns. The standard model for the infrared excess, and the mechanism that maintains the pollution, is the tidal disruption of asteroids and the circularisation of the debris into a disc \citep{Jura:2003aa}. As other authors have shown and we have discussed, the circularisation of the high eccentricity debris is inefficient. Therefore, we proposed that much or all of the debris may remain on highly eccentric orbits, producing the observed excess infrared flux by reprocessing flux emitted by the central white dwarf. Between the disruption and late times the debris spreads out to fill a highly eccentric ellipse; during this process the lightcurves emitted in different bands may exhibit large amplitude coherent variability, with the highest amplitude variability at short wavelengths (highest temperatures). The spectra and lightcurves we calculated are in reasonable agreement with the broad properties of the observations to date, including infrared spectral energy distributions and variability amplitude and timescales. We speculate that including additional physical processes may extend this simple model to account for additional observed phenomena; e.g. (1) perturbations to the debris stream from the planetary system---responsible for kicking in the original asteroid---may perturb the debris on to orbits with periastra inside the white dwarf surface, which would facilitate white dwarf pollution without the need for a dust/gas disc or significant excess infrared flux, and (2) that the self-gravity of the stream may play a significant role in determining the long-term evolution of the debris stream, including effects such as fragmentation of the stream and grinding the debris into small pieces. We will explore these processes in subsequent work.

\section*{Acknowledgments}
We are grateful to Laura Rogers for sharing a copy of Rogers et al. (submitted) with us prior to publication. CJN is supported by the Science and Technology Facilities Council (grant number ST/M005917/1), and funding from the European Union’s Horizon 2020 research and innovation program under the Marie Sk\l{}odowska-Curie grant agreement No 823823 (Dustbusters RISE project). ERC acknowledges support from NASA through the Hubble Fellowship, grant No. HST-HF2-51433.001-A awarded by the Space Telescope Science Institute, which is operated by the Association of Universities for Research in Astronomy, Incorporated, under NASA contract NAS5-265555. We used {\sc splash} \citep{Price:2007aa} for the figures. This work was performed using the DiRAC Data Intensive service at Leicester, operated by the University of Leicester IT Services, which forms part of the STFC DiRAC HPC Facility (\url{www.dirac.ac.uk}). The equipment was funded by BEIS capital funding via STFC capital grants ST/K000373/1 and ST/R002363/1 and STFC DiRAC Operations grant ST/R001014/1. DiRAC is part of the National e-Infrastructure.

\bibliographystyle{aasjournal}
\bibliography{nixon}

\end{document}